\documentclass[fleqn,twoside]{article}
\usepackage[headings]{espcrc2}

\readRCS
$Id: espcrc2.tex,v 1.2 2004/02/24 11:22:11 spepping Exp $
\usepackage{graphicx}
\usepackage[figuresright]{rotating}

\newcommand{\AmS}{{\protect\the\textfont2
  A\kern-.1667em\lower.5ex\hbox{M}\kern-.125emS}}

\title{Heavy Flavour Physics at CDF}

\author{Gavril Giurgiu on behalf of the CDF collaboration \\ 
	   Phyiscs and Astronomy Department,  
	   The Johns Hopkins University \\ 
	   3400 N. Charles Street, Baltimore, MD 21218, U.S.A.
        }
      
\runtitle{Heavy Flavour Physics at CDF}
\runauthor{G. Giurgiu}

\begin{document}

\begin{abstract}

The CDF detector at Fermilab has accumulated more that 3 fb$^{-1}$ of
data which enables unprecedented studies of heavy flavor hadron
properties. We present recent CDF measurements of mass and lifetime of the 
$B_c$ meson as well as lifetime, mixing and CP violation properties 
of $B_s$ mesons.

\vspace{1pc}
\end{abstract}

\maketitle

\section{Introduction}

We review recent measurements of $B_c$ and $B_s$ meson properties performed 
with the CDF detector in $\sqrt{s} = 1.96$~TeV $p \bar{p}$ collisions at the 
Tevatron. Due to the excellent performance of the Tevatron the CDF experiment 
has recorded 3.5~$fb^{-1}$ of data by June 2008. The measurements presented 
here use between 1~$fb^{-1}$ and 2.4~$fb^{-1}$. The CDF detector is described 
in detail in Ref.~\cite{cdf_detector}.

At the Tevatron, $B$ hadrons are mostly produced in pairs. The main $b \bar{b}$ 
production mechanism is flavor creation through gluon fusion.
The $b \bar{b}$ production 
cross section of $\sim30~\mu b$ \cite{cdf_detector} is large compared to B-factories 
which enables rich B Physics programs at the Tevatron. However, the Tevatron 
$b \bar{b}$ cross section is orders of magnitude smaller than the total 
inelastic cross section of $\sim50~mb$. For this reason, CDF employs triggers 
that select events with signatures specific to various B decays. The measurements 
presented here uses data selected by the di-muon trigger, geared towards $B \rightarrow J/\psi X$ 
decays and by the displaced track trigger used for measurement of $B_s$ lifetime 
in $B_s \rightarrow D_s \pi X$ decays.

\section{Mass and Lifetime of $B_c$ Mesons }

The $B_c$ meson is unique as it contains two heavy quarks: $b$ (bottom) and 
$\bar{c}$ (anti-charm). Different theoretical models predict the $B_c$ mass 
around $6.3~GeV/c^2$. In particular, non-relativistic potential models 
\cite{Bc_mass_potential} predict the range $6247-6286~MeV/c^2$.   
Lattice QCD models \cite{Bc_mass_NR_LQCD} predict $6304 \pm 12 ^{+18}_{-0}~MeV/c^2$, 
while similar results are determined in perturbative QCD calculations 
\cite{Bc_mass_pertQCD}. 

The lifetime of the $B_c$ meson is expected to be $\sim0.5~ps$ \cite{Bc_lifetime} 
which is about a third of the typical light B meson lifetime of $\sim1.5~ps$.
The short lifetime 
is explained by the fact that either the $b$ or the $c$ quarks can decay weakly 
through a $W$ boson. When the $b$ quark decays, the final state contains a charmonium 
particles while the decay of the $c$ quark results in a $B$ meson final state. 
Additionally, the $b$ and $c$ quarks can annihilate, for example into a lepton and neutrino final state.   
The measurement presented here use $B_c \rightarrow J/\psi\,l\,\nu\,X$ decays, where 
the lepton is either a muon or electron.

\subsection{Measurement of $B_c$ Mass}

\vskip0.2cm
The mass of the $B_c$ meson was measured for the first time by the CDF experiment in 
Run~I~\cite{Bc_mass_CDF_run_I} using inclusive semileptonic decays 
$B_c \rightarrow J/\psi\,l\,\nu\,X$. Since the 
$B_c$ meson was not fully reconstructed due to the missing neutrino, the result had 
large uncertainties $6400 \pm 390(stat.) \pm 130(syst.)~MeV/c^2$.
 
The most precise measurement of the $B_c$ mass was recently performed by the CDF 
experiment using $2.4~fb^{-1}$ of data \cite{Bc_mass_CDF_run_II} using the fully reconstructed decay mode 
$B_c \rightarrow J/\psi\,\pi$. The advantage of using fully reconstructed decays 
is that the mass of the decaying particle can be measured precisely by fitting 
the invariant mass distribution. The measurement was performed using 
data collected by the CDF $J/\psi$ trigger. This trigger requires two muons 
of opposite charge, transverse momentum larger than 1.4~GeV/c and invariant mass 
close to the $J/\psi$ mass. The data sample used in this analysis contains 
$\sim 17$~million $J/\psi$ candidates reconstructed with an average mass resolution 
of $13~MeV/c^2$. 

Using this initial $J/\psi$ sample, both $B^{+} \rightarrow J/\psi\,K^{+}$ and 
$B_c \rightarrow J/\psi\,\pi^{+}$ decays are reconstructed. The $B^{+}$ sample 
is used to determine an optimal set of selection requirements that are 
subsequently used for the $B_c$ selection. This procedure accounts for the 
expected shorter $B_c$ lifetime by using only $B^{+}$ mesons with decay 
time in rages typical to expected $B_c$ lifetime: $80 < ct < 300\,\mu m$.      

\begin{figure}[htb]
\vspace{9pt}
\includegraphics[width=70mm]{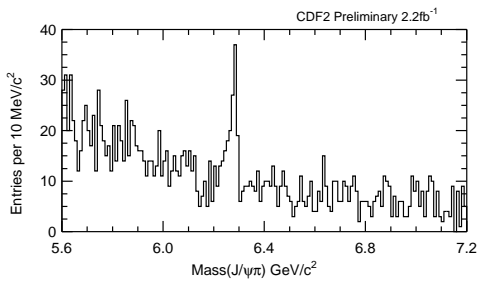}
\caption{$B_c$ invariant mass distribution reconstructed from 
	$B_c \rightarrow J/\psi~\pi$ decays.}
\label{fig:bc_mass1}
\end{figure}

The invariant mass distribution of the $B_c$ candidates is shown in 
Figure
~\ref{fig:bc_mass1}. The mass of the $B_c$ 
meson is determined using an un-binned log likelihood method. The signal 
probability density function (PDF) is described by a Gaussian function with 
a width proportional to the estimated mass uncertainty estimated from 
track uncertainties. The background PDF includes both a combinatorial 
component as well as a component from $B_c \rightarrow J/\psi\,K^{+}$ 
Cabibbo suppressed decays. 

The total $B_c$ signal yield is estimated to $108 \pm 15$ candidates 
with a statistical significance of 8 standard deviations. The measured mass 
is $6275.6 \pm 2.9~MeV/c^{2}$. The largest systematic uncertainty of $2.2~MeV/c^{2}$ 
is due to the knowledge of the scale factor on mass uncertainties. The total   
systematic uncertainty which includes effects due to misalignment, momentum 
scale and Cabibbo suppressed $B_c \rightarrow J/\psi\,K^{+}$ decays is estimated 
to $2.5~MeV/c^{2}$, comparable but slightly lower than the statistical error, 
leading to the most precise measurement of the $B_c$ mass: 
$6275.6 \pm 2.9(stat.) \pm 2.5(syst.)~MeV/c^{2}$, in good agreement with the corresponding 
D0 measurement \cite{Bc_mass_D0} $6300 \pm 14(stat.) \pm 5(syst.)~MeV/c^{2}$

\subsection{Measurement of $B_c$ Lifetime}

\vskip0.2cm
The lifetime of the $B_c$ meson is measured by the CDF experiment using partially 
reconstructed semileptonic $B_c \rightarrow J/\psi\,(\mu\mu)\,l\,\nu\,X$ decays \cite{Bc_lifetime_CDF} 
in $1.0~fb^{-1}$ of data. The lepton $l$ can be either a muon or an electron. The data 
sample is selected by the same $J/\psi$ trigger used for the $B_c$ mass measurement.   
The lifetime analysis starts with a sample of $\sim 5.5$ million $J/\psi \rightarrow \mu^{+}\,\mu^{-}$ 
candidates. A third lepton is then required to come from the same vertex as $J/\psi$. 
The main challenges in this measurement are the partially reconstructed momentum of the 
$B_c$ meson and multiple backgrounds. 

Since the momentum of the $B_c$ meson is not 
fully reconstructed, the observed pseudo proper decay time $ct^{*} = \frac{m\,L_{xy}(J/\psi\, l)}{p_T(J/\psi\, l)}$ 
is statistically corrected by a factor $K = \frac{p_T(J/\psi\, l)}{p_T(B_c)}$ determined 
from simulation. Here, $m$ is the mass of the $B_c$ meson \cite{Bc_mass_CDF_run_II_first}, 
$p_T(J/\psi\,l)$ is the reconstructed transverse momentum of the $J/\psi$ and lepton,  
$p_T(B_c)$ is the true transverse momentum of the $B_c$ meson determined from simulation and   
$L_{xy}$ is the transverse $B_c$ decay length projected on the reconstructed 
transverse momentum.   

In this measurement the mass of the $B_c$ meson is not reconstructed 
due to the missing neutrino. Instead, the $J/\psi\,l$ mass is used. However, 
this quantity has a wide distribution and sidebands cannot be used to estimate 
backgrounds as in fully reconstructed decays. Different background contributions 
are determined using data-driven methods when possible and simulation 
otherwise. The main sources of backgrounds are: fake leptons, fake $J/\psi$, 
``$b \bar{b}$ sources'', $e^{+}e^{-}$ from either photon conversions or decays 
of $\pi^{0}$ and $\eta$ mesons and prompt $J/\psi$ events.    
The fake lepton background is produced by a real $J/\psi$ and a random track 
which fakes a muon or an electron and forms a $J/\psi\,l$ vertex that satisfies 
the vertex quality requirements. The fake lepton fractions are determined 
using real pions and kaons from $D^{*+} \rightarrow D^{0}\, \pi^{+}$ with 
$D^0 \rightarrow \pi^{+} \, K^{-}$ and real protons from $\Lambda \rightarrow p^{+} \, \pi^{-}$, 
where the real pions, kaons and protons are misidentified as leptons. 
The fake $J/\psi$ background is modeled using the $J/\psi$ mass sidebands. 
$b \bar{b}$ backgrounds are events with a real $J/\psi$ from one $b$ quark and 
a real lepton from the other $\bar{b}$ anti-quark in the event. This background is 
modeled using Pythia \cite{Pythia} simulation where the fractions of the three 
main production mechanisms, flavor creation, flavor excitation and 
gluon splitting are tuned to data using the angular separation between 
the $J/\psi$ and the lepton as discriminating variable.  
$e^{+}e^{-}$ backgrounds, specific to $J/\psi~e$ mode only, are suppressed 
by trying to identify both the positron and electron from photon conversion 
or light meson decay. The residual $e^{+}e^{-}$ pairs are modeled 
using Pythia \cite{Pythia} simulation. 
The background fraction from prompt $J/\psi$ is modeled by a Gaussian function 
in the pseudo proper decay space and determined directly from the un-binned 
likelihood fit.  

\begin{figure}[htb]
\vspace{9pt}
\includegraphics[width=70mm]{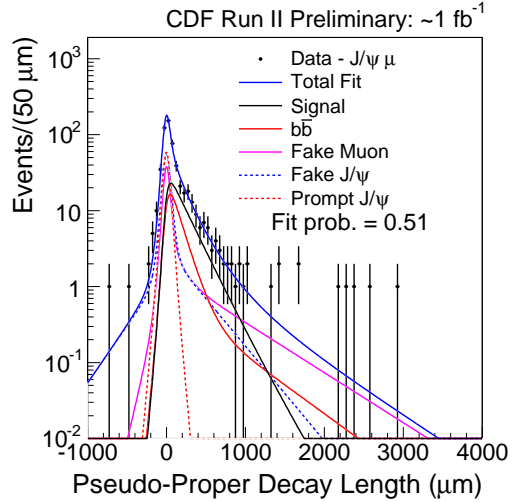}
\caption{ Pseudo proper decay time fit projections in $B_c \rightarrow J/\psi\,\mu\,\nu\,X$ channel. }
\label{fig:bc_lifetime_mu}
\end{figure}

The $B_c$ lifetime is measured separately in $J/\psi\,e$ and $J/\psi\,\mu$ 
channels using un-binned likelihood methods. The fit projections of the pseudo proper 
decay time distributions in the $J/\psi~\mu$ channel is shown in Figure~\ref{fig:bc_lifetime_mu}. 

\begin{figure}[htb]
\vspace{9pt}
\includegraphics[width=70mm]{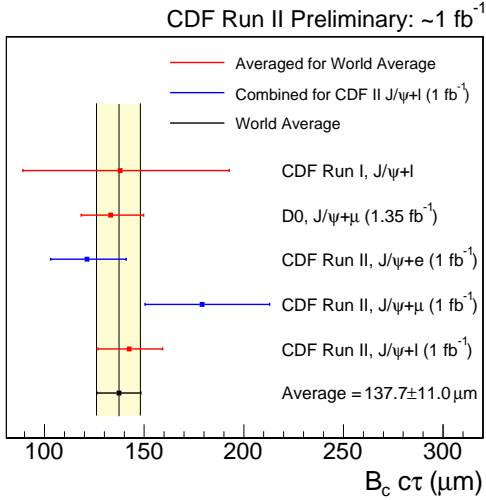}
\caption{ Summary of the $B_c$ lifetime experimental results.}
\label{fig:bc_summary}
\end{figure}

The corresponding lifetimes are: $c\tau_{\mu} = 179.1 ^{+32.6}_{-27.2}(stat.)$ 
and $c\tau_{e} = 121.7 ^{+18.0}_{-16.3}(stat.)\,\mu m$      
The main sources of systematic uncertainties come from our understanding of the 
decay time resolution function ($3.8\,\mu m$) and from relative fractions of $b \bar{b}$ 
production mechanisms in Pythia simulation. The total systematic uncertainty is $5.5\,\mu m$, 
leading to a combined measurement of the $B_c$ lifetime of 
$c\tau_{\mu} = 142.5 ^{+15.8}_{-14.8}(stat.) \pm 5.5(syst.)\,\mu m$. 
This agrees well with a similar recent measurement performed by the D0 experiment \cite{Bc_lifetime_D0} 
in the $B_c \rightarrow J/\psi \mu~\nu~X$  channel $c\tau = 134.4 ^{+11.4}_{-10.8}(stat) \pm 9.6(syst)$. 
 Figure~\ref{fig:bc_summary} shows a summary of the experimental results of the 
$B_c$ lifetime. CDF and D0 measurements agree well with the theoretical 
predictions and with each other and have similar uncertainties. 

\section {Neutral $B_s$ System} 

A $B_s^0$ meson is a bound state composed of an anti-bottom quark $\bar {b}$ 
and a strange $s$ quark. The time evolution of a mixture of the $B_s^0$ and 
its antiparticle $\bar {B}_s^0$, $a(t)|B^0_s\rangle+b(t)|\bar{B}^0_s\rangle$, 
is given by the Schr$\ddot{{\rm o}}$dinger equation
\begin{equation}
i\frac{d}{dt} \left(\begin{array}{c}
a \\
b \end{array}\right) = \left(M-i\frac{\Gamma}{2}\right) \left(\begin{array}{c}
a \\
b \end{array}\right),
\label{eqn:Schrodinger}
\end{equation}
where $M$ and $\Gamma$ are $2\times2$ mass and decay matrices. 
The mass eigenstates $B^0_L$ and $B^0_H$ are liner combination 
of the flavor eigenstates $B_s^{0}$ and $\bar{B}_s^0$ and are obtained by 
diagonalizing the mass operator. 
The mass difference $\Delta m_s$ is proportional 
to the $B_s$ mixing frequency recently measured by both CDF \cite{dms_cdf} 
and D0 \cite{dms_d0} experiments and found to be in good agreement with 
Standard Mmodel (SM) predictions \cite{Ref:lenz}. Moreover, the mass eigenstates have different 
decay widths (lifetimes) $\Gamma_L$ and $\Gamma_H$. The average 
decay width is defined as $\Gamma=(\Gamma_L+\Gamma_H)/2$ and the decay width 
difference is defined as $\Delta\Gamma=\Gamma_L-\Gamma_H$.
The decay width difference $\Delta\Gamma=2|\Gamma_{12}|\cos(\phi_s)$ 
is sensitive to new physics (NP) effects~\cite{Ref:hou,Ref:ligeti,Ref:lenz}    
that affect the phase $\phi_s = {\rm arg}(-M_{12}/\Gamma_{12})$, where $\Gamma_{12}$ and $M_{12}$ 
are the off-diagonal elements of the mass and decay matrices. New physics will  
increase $\phi_s$, so $\Delta\Gamma$ would be smaller than the SM 
prediction. Since the SM phase $\phi_s^{SM}$ is predicted to be very 
small ($\sim 0.004$) \cite{Ref:lenz}, in a new physics scenario with large 
contribution to $\phi_s$ one could approximate $\phi_s = \phi_s^{SM} + \phi_s^{NP} \approx 
\phi_s^{NP}$. If real, this new physics phase is accessible in $B_s \rightarrow J/\psi\,\phi$ 
decays. In these decays one can measure the CP violation phase $\beta_s$ which 
is the relative phase between the direct decay amplitude and mixing followed by decay amplitude. 
In SM this phase is defined as        
$\beta_s^{\textit{SM}} = \arg\left(-\frac{V_{ts}V_{tb}^*}
{V_{cs}V_{cb}^*}\right)$~\cite{Ref:lenz}, where $V_{ij}$ are the
elements of the CKM quark mixing matrix. Global fits of experimental
data tightly constrain the CP~violation phase to small values in the
context of SM $\beta_s^{SM}\approx 0.02$~\cite{Ref:HFAG}. The presence of new physics 
could modify this phase by the same quantity $\phi_s^{NP}$ that affects the $\phi_s^{SM}$ phase. 
New physics could contribute significantly to the observed
$\beta_s$~phase~\cite{Ref:hou,Ref:ligeti,Ref:lenz} expressed as 
$2\beta_s = 2\beta_s^{\textit{SM}} - \phi_s^{\textit{NP}}$. Assuming 
that new physics effects dominate over the SM phase, 
we can approximate $2\beta_s \approx - \phi_s^{NP} \approx -\phi_s$.   

\subsection{$B_s$ Lifetime and Width Difference}

\begin{figure}[htb]
\vspace{9pt}
\includegraphics[width=70mm]{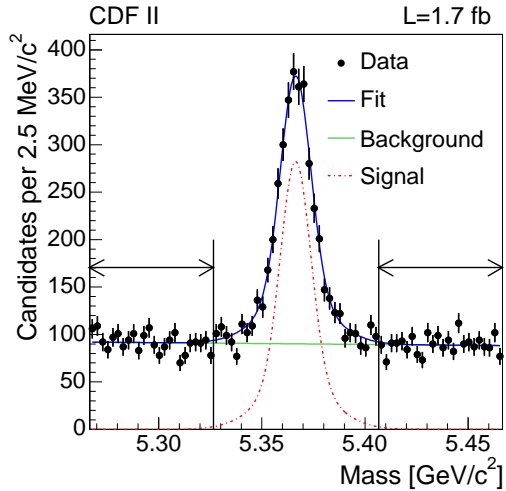}
\caption{ $B_s$ invariant mass from $B_s \rightarrow J/\psi\,\phi$ decays. 
	Fit projections are superimposed. }
\label{fig:bs_mass}
\end{figure}

\vskip0.2cm
A sample of $\sim 2500$ signal $B_s \rightarrow J/\psi\,\phi$ events 
collected with the CDF $J/\psi$ trigger in $1.7~fb^{-1}$ of data was 
used to measure the $B_s$ lifetime $\tau_s$ and decay with difference 
$\Delta \Gamma_s$ \cite{Ref:Kuhr}. The $B_s$ invariant mass distribution is seen in 
Figure~\ref{fig:bs_mass} which shows good signal to background ratio. 
The $B_s$ meson is a spin 0 particle which decays in two spin 1 particles 
$J/\psi$ and $\phi$. The total angular momentum in the final state can be 0, 1 
or 2. The states with angular momentum 0 and 2 are CP even, while 
the state with angular momentum 1 is CP odd. One can statistically separate 
the CP even and CP odd states using their different angular distributions. 
In the case of CP conservation the mass eigenstates are also CP eigenstates. 
It is interesting to note that since in SM the CP violation phase 
in the $B_s$ system is very small, the mass eigenstates are CP eigenstates 
with good approximation.     

\begin{figure}[htb]
\vspace{9pt}
\includegraphics[width=70mm]{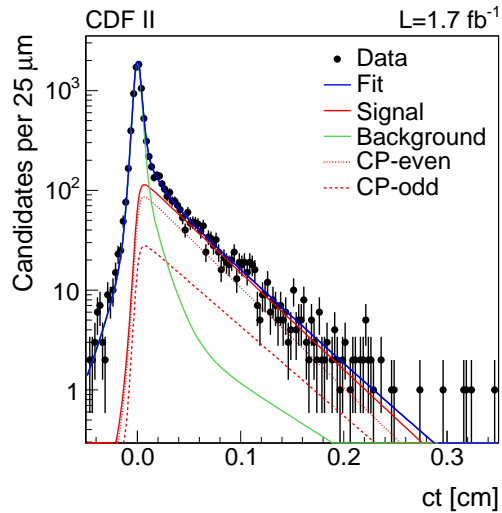}
\caption{ $B_s$ decay time from $B_s \rightarrow J/\psi\,\phi$ decays. 
	Fit projections are superimposed. }
\label{fig:bs_lifetime}
\end{figure}

The $B_s$ lifetime and decay width difference are extracted using an un-binned 
likelihood method in the space of mass, decay time and angles of the final state 
particles. There are three angles $\vec\rho = \{\cos\theta_T,\phi_T, \cos\psi_T\}$
that completely define the directions of the 
four particles $\mu^{+}, \mu^{-}, K^{+}, K^{-}$ in the final state. These 
angles are defined in the transversity basis introduced in Ref.~\cite{Ref:Dighe}. 
Figure~\ref{fig:bs_lifetime} shows the $B_s$ decay time distribution together 
with the fit projection. The largest systematic uncertainties on lifetime are due 
to decay time resolution model and alignment of the silicon detector. The 
largest systematic on $\Delta \Gamma$ is due to about $3\%$ of 
$B^0 \rightarrow J/\psi\, K^{*0}$ decays reconstructed as $B_s \rightarrow J/\psi\, \phi$. 
The final results are: $c\tau = 1.52 \pm 0.04(stat.) \pm 0.02(syst.)~ps$
and $\Delta \Gamma = 0.076 ^{+0.059}_{-0.063}(stat.) \pm 0.006(syst.) ps^{-1}$.
The fractions of each of the three final state amplitudes are also measured 
and together with the lifetime and width difference are found to be in agreement 
with the similar measurement performed by the D0 experiment \cite{untagged_D0}.
    
\subsection{CP Violation in $B_s$ System}

\vskip0.2cm
The CP violation phase $\beta_s$ in $B_s \rightarrow J/\psi\, \phi$ decays 
was measured for the first time by the CDF experiment using $1.4~fb^{-1}$ 
of data \cite{ref:tagged_cdf}. The analysis uses 
a sample of $\sim 2000$ signal events. To enhance our sensitivity to 
the CP violation phase, we identify the flavor of the $B_s$ or
$\bar {B}_s$~meson at production by means of flavor tagging.
Two independent types of flavor tags are used, each exploiting
specific features of the production of $b$~quarks at the Tevatron. 
The first type of flavor tag infers the production flavor of the $B_s$ or 
$\bar{B}_s$~meson from the decay products of the $b$~hadron produced by the other
$b$~quark in the event.  This is known as an opposite-side flavor tag
(OST). 
The second type of flavor tag identifies the
flavor of the reconstructed $B_s$ or $\bar{B}_s$~meson at production by
correlating it with the charge of an associated kaon arising from
fragmentation processes, 
referred to as a same-side
kaon tag (SSKT). 
The average dilution $D$, defined via the correct tag probability $P = (1+D)/2$,  
is $(11\pm 2)\%$ for the OST and $(27\pm 4)\%$ for the SSKT. 
The measured efficiencies for a candidate to be tagged are
$(96\pm 1)\%$ for the OST and $(50 \pm 1)\%$ for the SSKT.

An unbinned maximum likelihood fit is performed to extract the
parameters of interest, $2\beta_s$ and $\Delta\Gamma$, plus additional
parameters referred to as ``nuisance parameters'' which include the
signal fraction 
$f_s$, the mean $B_s$ width $\Gamma\equiv (\Gamma_L + \Gamma_H)/2$,
the mixing frequency $\Delta m_s$, the magnitudes of the polarization
amplitudes $|A_0|^2$, $|A_{\parallel}|^2$, and $|A_{\perp}|^2$, and
the strong phases $\delta_{\parallel}\equiv
\arg(A^*_{\parallel}A_0)$ and $\delta_{\perp}\equiv
\arg(A^*_{\perp}A_0)$. The fit uses information on the
reconstructed $B_s$ candidate mass $m$ and its uncertainty $\sigma_m$,
the $B_s$ candidate proper decay time $t$ and its uncertainty
$\sigma_{t}$, the transversity angles $\vec\rho$
and tag information ${\cal D}$ and $\xi$, where
${\cal D}$ is the event-specific dilution and $\xi=\{-1,0,+1\}$ is the
tag decision, in which $+1$ corresponds to a candidate tagged as
$B_s$, $-1$ to a $\bar{B}_s$, and $0$ to an untagged candidate.  

%


The time and angular dependence of the signal PDF 
$P_s(t,\vec\rho,\xi,|{\cal D},\sigma_{t})$
can be written in terms of two PDFs, $P$ for $B_s$ and $\bar{P}$ for
$\bar{B}_s$, as
\begin{eqnarray}
  P_s(t,\vec\rho,\xi|{\cal D},\sigma_{t}) 
      &=& \frac{1+\xi{\cal D}}{2} P(t,\vec\rho| \sigma_{t}) \epsilon(\vec\rho)  \nonumber \\
      &&+\frac{1-\xi{\cal D}}{2}\bar{P}(t, \vec\rho| \sigma_{t}) \epsilon(\vec\rho),
\label{eqn:signalPDF}
\end{eqnarray}
 The detector acceptance effects on the transversity
angle distributions $\epsilon(\vec\rho)$ are modeled with
$B_s \rightarrow J/\psi\, \phi$ simulated data.  
The time and angular probabilities for $B_s$ can be expressed as
\begin{eqnarray}
P(t,\vec\rho)
&\propto& |A_0|^2 {\cal T}_+ f_1(\vec\rho) + |A_\parallel|^2 {\cal T}_+ f_2(\vec\rho) \nonumber \\
&+& |A_\perp|^2 {\cal T}_- f_3(\vec\rho) + |A_\parallel||A_\perp| {\cal U}_+ f_4(\vec\rho) \nonumber \\
&+& |A_0||A_\parallel| \cos(\delta_{\parallel}) {\cal T}_+ f_5(\vec\rho) \nonumber \\
&+& |A_0||A_\perp| {\cal V}_+ f_6(\vec\rho), \label{eqn:Bs} 
\end{eqnarray}
where the functions $f_1(\vec\rho)\dots f_6(\vec\rho)$ are defined in
Ref.~\cite{Ref:Kuhr}.  The probability $\bar{P}$ for $\bar{B}_s$ is
obtained by substituting ${\cal U_+}\rightarrow{\cal U_-}$ and ${\cal
V_+}\rightarrow{\cal V_-}$.  The time-dependent term ${\cal T}_{\pm}$ is
defined as
\begin{eqnarray*}
{\cal T}_{\pm}=e^{-\Gamma t}&\times&\left[\cosh(\Delta \Gamma t/2)
                 \mp \cos(2\beta_s)\sinh(\Delta \Gamma t/2)\right.\\
              & &\left.\mp\ \eta\sin(2\beta_s)\sin(\Delta m_st)\right], 
\end{eqnarray*}
where $\eta = +1$ for $P$ and $-1$ for $\bar{P}$.  The other
time-dependent terms are defined as
\begin{eqnarray*}
{\cal U}_{\pm} =\pm e^{-\Gamma t}&\times&\left[\sin(\delta_{\perp}-\delta_{\parallel})\cos(\Delta m_st) \right.\\
               &-&\cos(\delta_{\perp}-\delta_{\parallel})\cos(2\beta_s)\sin(\Delta m_st)                \\
               &\pm&\left.\cos(\delta_{\perp}-\delta_{\parallel})\sin(2\beta_s)\sinh(\Delta\Gamma t/2)
                  \right], \nonumber\\
{\cal V}_{\pm} =\pm e^{-\Gamma t}&\times&\left[\sin(\delta_{\perp})\cos(\Delta m_st) \right.\\
               &-&\cos(\delta_{\perp})\cos(2\beta_s)\sin(\Delta m_st)  \\
               &\pm&\left.\cos(\delta_{\perp})\sin(2\beta_s)\sinh(\Delta\Gamma t/2)\right].
\end{eqnarray*}
The time-dependence is convolved with a Gaussian
proper time resolution function with standard deviation $\sigma_{t}$,
which is adjusted by an overall calibration factor determined from the
fit using promptly decaying background candidates.  
The average of the resolution function is 0.09~ps, with a root-mean-square 
deviation of 0.04~ps.


Possible asymmetries between the tagging rate and dilution of $B_s$
and $\bar{B}_s$~mesons have been studied with control samples and found to
be statistically insignificant.  We allow important sources of
systematic uncertainty, such as the determination of overall
calibration factors associated with the proper decay time resolution
and the dilutions, to float in the fit.  The mixing frequency $\Delta
m_s = 17.77 \pm 0.12$~ps$^{-1}$ is constrained in the fit within the
experimental uncertainties~\cite{dms_cdf}. Systematic
uncertainties coming from alignment, detector sculpting, background
angular distributions, decays from other $B$~mesons, the modeling of
signal and background are found to have a negligible effect on the
determination of both $\Delta\Gamma$ and $\beta_s$ relative to
statistical uncertainties.

An exact symmetry is present in the signal probability distribution,
as can be seen in Eq.~(\ref{eqn:Bs}), which is invariant under the
simultaneous transformation ($2\beta_s\to\pi-2\beta_s$,
$\Delta\Gamma\to -\Delta\Gamma$, $\delta_{\parallel}\to 2\pi -
\delta_{\parallel}$, and $\delta_{\perp}\to \pi -\delta_{\perp}$).
This causes the likelihood function to have two minima.  
Since the log-likelihood function is non-parabolic and the two minima 
are barely separated at one standard deviation level, we cannot
meaningfully quote point estimates. Instead we choose to construct a
confidence region in the $2\beta_s-\Delta\Gamma$ plane.

\begin{figure}[htb]
\includegraphics[width=70mm]{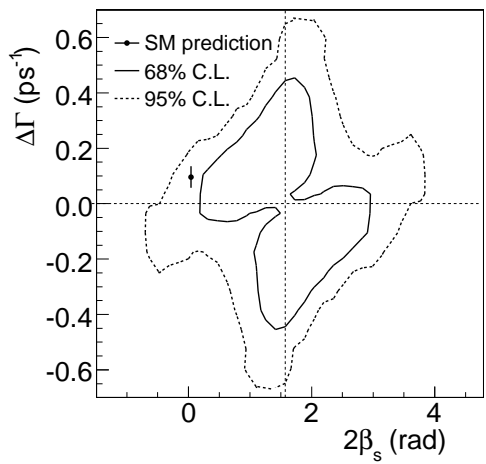}
\caption{ Confidence region in the $2\beta_s-\Delta\Gamma$
	plane, where the standard model favored point is shown with error
	bars~\cite{Ref:lenz}. }
\label{fig:contour}
\end{figure}

We use the Feldman-Cousins likelihood ratio ordering to
determine confidence levels (CL) in $2\beta_s-\Delta\Gamma$ space. 
The other parameters in the fit are treated
as nuisance parameters ({\it e.g.} $B_s$ mean width, transversity
amplitudes, strong phases). To ensure that the obtained
confidence regions provide the quoted coverage against deviations of the
nuisance parameters from their values measured in our fit to data, we perform
pseudo-experiments by randomly sampling the nuisance parameter space within
$\pm 5\sigma$ of the fit values and confirm
coverage of the 68\% and 95\% confidence regions shown in
Fig.~\ref{fig:contour}.  The solution centered in $0\leq 2\beta_s
\leq\pi/2$ and $\Delta\Gamma > 0$ corresponds to $\cos(\delta_{\perp})
< 0$ and $\cos(\delta_{\perp}-\delta_{\parallel}) > 0$, while the
opposite is true for the solution centered in $\pi/2 \leq \beta_s \leq
\pi$ and $\Delta\Gamma <0$.  Assuming the standard model predicted
values of $2\beta_s=0.04$ and $\Delta\Gamma = 0.096\mbox{
ps}^{-1}$~\cite{Ref:lenz}, the probability to observe a likelihood
ratio equal to or higher than what is observed in data is 15\%.
Additionally, we present a Feldman-Cousins confidence interval of
$2\beta_s$, where $\Delta\Gamma$ is treated as a nuisance parameter,
and find that $2\beta_s\in[0.32,2.82]$ at the 68\% confidence level.
The CP~phase $2\beta_s$, $\Delta\Gamma$, $\Gamma$, and the
linear polarization amplitudes are consistent with those measured in
Ref.~\cite{Ref:Kuhr} and in Ref.~\cite{untagged_D0}.

\subsection{$B_s$ Lifetime in Flavor Specific Decays}

\begin{figure}[htb]
\includegraphics[width=70mm]{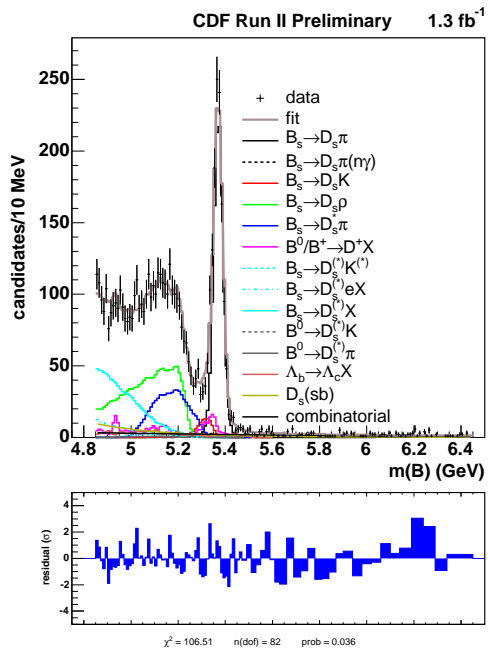}
\caption{ Invariant $B_s$ mass distribution and fit projection. }
\label{fig:bs_mass_flvsp}
\end{figure}

\vskip0.2cm
The lifetime of the $B_s$ meson is measured by the CDF collaboration 
in flavor specific $B_s \rightarrow D_s\, \pi \, X$ decays \cite{ref:bs_lifetime_flvsp} 
using $1.3~fb^{-1}$ of data. The flavor of the $B_s$ meson at decay is given by 
the charges of the final state particles, hence the name ``flavor specific''. 
These events are collected by a trigger which requires 
two tracks with impact parameter between $120\,\mu m$ and $1~mm$, specific to 
hadronic B decays. The sample contains $\sim 1100$ fully reconstructed 
$B_s \rightarrow D_s (\phi\,\pi)\, \pi$ and $\sim 2200$ partially reconstructed decays.
Figure~\ref{fig:bs_mass_flvsp} shows the $B_s$ invariant mass distribution 
and details of all contributions to partially reconstructed decays. 

There are two main challenges in this analysis. First, the lifetime distribution of 
the $B_s$ meson is biased due to the trigger that requires tracks with 
large impact parameter. This is corrected by using an efficiency function determined 
from simulation. This method is validated on control samples of $B^0 \rightarrow D^{-} \pi^{+}$, 
$B^0 \rightarrow D^{*-} \pi^{+}$ and $B^+ \rightarrow \bar{D}^{-} \pi^{+}$ decays 
where good agreement with the $B^0$ and $B^{+}$ world average lifetimes is obtained. 
A second challenge is to determine background models from the upper $B_s$ mass sideband 
as well as from wrong sign $B_s \rightarrow D_s^{+}\, \pi^{+}$ events. 
Additionally, we determine from simulation the mass distributions and momentum 
correction of partially reconstructed modes. The main systematic uncertainties 
come from background modeling. The measured lifetime is 
$c\tau_{s} = 455.0 \pm 12.2(stat.) \pm 7.4(syst.)\,\mu m$. This is the best 
measurement in flavor specific decays and agrees very well with both CDF and D0 
measurements in $B_s \rightarrow J/\psi\, \phi$ decays discussed in section 3.1.     
We note that the lifetime measured in flavor specific decays is not exactly 
the inverse of the decay width as measured in $B_s \rightarrow J/\psi\, \phi$ decays. 
The actual relation between the measured quantity in flavor specific decays 
and the actual lifetime is: 
$\tau(B^0_s)_{FS}= \frac{1}{\Gamma}(1+(\frac{\Delta\Gamma}{2\Gamma})^2)/(1-(\frac{\Delta\Gamma}{2\Gamma})^2)$
which includes a small correction which depends on the width difference $\Delta \Gamma$. 
It is also interesting to note that this measurement together with 
the lifetime measurements in $B_s \rightarrow J/\psi\, \phi$ decays 
from both CDF~\cite{Ref:Kuhr} and D0 \cite{untagged_D0} are larger than the 2007 
world average~\cite{Ref:HFAG} ($1.41 \pm 0.04~ps$). The new world average $B_s$ lifetime will be 
much closer to the $B^0$ lifetime as predicted theoretically.

\section{Summary}

We have presented recent CDF results on $B_c$ and $B_s$ meson properties. 
The best $B_c$ mass measurement was performed by CDF in fully reconstructed 
$B_c \rightarrow J/\psi\, \pi$ decays. Measurements of the $B_c$ lifetime 
in semileptonic decays agrees with the measurement from the D0 
experiment and has similar precision. These measurements provide useful 
information that can be used to improve theoretical models used to study 
heavy mesons. 
The recent measurements of the $B_s$ lifetime in both $B_s \rightarrow J/\psi\,\phi$ 
and flavor specific $B_s \rightarrow D_s\, \pi$ channels confirm the 
theoretical predictions that $\tau_s \approx \tau_d$. The measurements  
of the decay width difference $\Delta \Gamma$ and the CP violation 
phase $\beta_s$ are statistically limited but open the road for exciting 
results with full Tevatron data sets.

\vskip0.2cm
I would like to thank the organizers of BEACH 2008 conference for an enjoyable experience 
and my CDF colleagues and Fermilab staff for their dedication which 
makes these results possible.

\end{document}